\documentclass[conference]{IEEEtran}
\IEEEoverridecommandlockouts
\usepackage{cite}
\usepackage{amsmath,amssymb,amsfonts}
\usepackage{algorithmic}
\usepackage{graphicx}
\usepackage{textcomp}
\usepackage{xcolor}
\usepackage{float}
\usepackage{enumitem}
\def\BibTeX{{\rm B\kern-.05em{\sc i\kern-.025em b}\kern-.08em
    T\kern-.1667em\lower.7ex\hbox{E}\kern-.125emX}}
\begin{document}

\title{Fault-Tolerant IoT System Using Software-Based ``Digital Twin”\\
{\footnotesize \textsuperscript{*}}
}

\makeatletter
\newcommand{\linebreakand}{%
  \end{@IEEEauthorhalign}
  \hfill\mbox{}\par
  \mbox{}\hfill\begin{@IEEEauthorhalign}
}
\makeatother

\author{\IEEEauthorblockN{Tanish Baranwal}
\IEEEauthorblockA{\textit{DomaniSystems, Inc.} \\
Shelton, CT 06484 \\
b.tanish@gmail.com}
\and
\IEEEauthorblockN{Srihari Varada}
\IEEEauthorblockA{\textit{DomaniSystems, Inc.} \\
Shelton, CT 06484 \\
sriharavarada@hotmail.com} \\
\and
\IEEEauthorblockN{Santanu Das}
\IEEEauthorblockA{\textit{DomaniSystems, Inc.} \\
Shelton, CT 06484 \\
das@domanisystems.com}
\linebreakand 
\IEEEauthorblockN{Mohammad R. Haider}
\IEEEauthorblockA{\textit{University of Missouri} \\
\textit{Department of Electrical Engineering and Computer Science}\\
Columbia, MO 65211 \\
mhaider@missouri.edu}
}

\maketitle

\begin{abstract}
In this article, we present a novel redundancy scheme to realize a fault-tolerant IoT structure for application in high-reliability systems. The proposed fault-tolerant structure uses a centralized data fusion block and triplicated IoT devices, along with software-based ``digital twins”, that duplicate the function of each of the sensors. In case of a fault in one of the IoT devices, the pertinent digital twin takes over the function of the actual IoT device for some time in the triplicated structure till the faulty device is either replaced or repaired when possible. The use of software-based digital twins as a duplicate for each physical sensor improves the reliability of the operation with minimal increase in the overall system cost. 
\end{abstract}


\section{Introduction}
The Internet of Things (IoT) generally refers to devices with sensors, processing ability, software, and other technologies that connect and exchange data with other devices and systems over the Internet or other communications networks [1].
Real-world Internet of Things examples range from a smart home that automatically adjusts heating and lighting to a smart factory that monitors industrial machines to look for problems, and then automatically adjusts to avoid failures [2].
The presence of faults in IoT systems can cause economic loss and even fatal accidents in applications like Autonomous Electric Vehicles (EVs), undermining the trustworthiness of IoT systems in the public eye [3, 4].
A fault in a sensor network is an unexpected behavior of a sensor node that leads to network failure. Faults can arise due to numerous external and internal sources in the environment, such as the deployment of sensor nodes in a hostile environment, the hardware or software malfunction in the sensor device, and the occurrence of natural disasters [5, 6].
The abnormal or unexpected behavior of a sensor node is called a fault in the sensor [5, 6]. The fault in active mode deviates from the expected results of the designed model and gives erroneous outcomes. {\em The sensor node in a sensor network either does not respond to its environment (hard fault) or responds with erroneous results every time (soft fault) or sometimes behaves fault-free and sometimes faulty (intermittent fault), or abnormal behavior persists for a short period and then vanishes suddenly (transient fault).}
Faults in a sensor network can be dealt with by implementing fault-tolerant structures. These structures generally contain three fundamental capabilities: error detection, error diagnosis, and error recovery [7].
There are two main approaches to detecting, identifying, and masking sensor faults to ensure that the system performs the required tasks as expected. These are hardware redundancy and analytical redundancy [8]. The first approach uses the fact that several sensors measure the same quantity. The second approach utilizes a mathematical model of the system, for example, a finite element model, and the redundancy is provided by the model. 
The use of redundancy to improve fault tolerance involves deploying duplicate or backup components to ensure uninterrupted operations in case of failures. Redundancy can be applied at various levels, including hardware, software, and data. For example, redundant sensors or actuators can be installed to maintain functionality even if some devices fail. Redundant data storage and backup systems can ensure data integrity and availability [9].
An example of the use of hardware redundancy to realize a fault-tolerant IoT system would be to use multiple sensors or IoT devices to monitor the same parameter in high-reliability applications like an industrial machine. The types of items that are monitored in an industrial machine include leaks, vibration, emissions, data, pressure, voltage, etc.
An example of the use of analytical redundancy is in the monitoring of traffic in a smart city. In this case, multiple sensors are installed at different locations to measure the traffic flow. The readings from these sensors are then compared to detect any discrepancies. If a discrepancy is detected, the faulty sensor can be identified and replaced, ensuring that the traffic data remains accurate and reliable.
Ideally, the fault-tolerant IoT Systems should be able to deal with different types of fault conditions including hard fault, soft fault, intermittent fault, and transient fault.
Curiac et al. [10] provide a comprehensive survey of various approaches for the use of redundancy in wireless sensor networks, related to sensing, communication, and information processing. They also present two methodologies: one that implies the use of components for spatial redundancy and the other that implies the use of temporal redundancy for achieving the objective of fault-tolerant and safe operation.
One of the major challenges in using redundant sensors to construct fault-tolerant IoT networks is to maintain a long network lifetime as well as sufficient sensing area. To achieve this goal, a broadly-used method is to turn off redundant sensors. In [11], the problem of estimating redundant sensing areas among neighboring wireless sensors is analyzed.
The objective here is to ensure that Wireless Sensor Networks (WSNs) can operate for a long time with little or no external management. The sensor network must be able to reconfigure itself in the presence of adverse situations [12, 13]. 

One of the main weaknesses of using hardware redundancy is that it can be costly. Another weakness is that hardware redundancy can increase power consumption. 
The main weakness of analytical redundancy is that it can be computationally expensive and may require a lot of memory. Another weakness is that it may not be effective in detecting faults that occur simultaneously in multiple sensors [14].

In this article, we present a novel redundancy scheme to realize a fault-tolerant IoT structure for application in high-reliability systems like industrial machinery, Electric Vehicles, Airplane engines, etc. where the mean time to repair is measured in hours or sometimes in days. The proposed fault-tolerant structure uses triplicated sensors along with software-based “digital twins” [15] that duplicate the function of each of the sensors. This triplicated structure is based on the concept of triple modular redundancy (TMR) [16], where three IoT devices process the same parameter, like reading the temperature of an oven, and that result is processed by a majority-voting system to produce a single output. If any one of the three sensors malfunctions, the other two systems can correct and mask the fault. This is explained in more detail in the next Section. 
A digital twin is a virtual model designed to accurately reflect a physical object [15]. The digital twin embeds extensive data regarding the behavior of the physical device under different conditions to enable the digital twin to mimic the behavior of the physical device. Having the digital twin embedded in the TMR structure ensures the integrity of the majority voting scheme and improves the overall reliability of the system. This is because the digital twin mimics the operation of the sensor, which malfunctioned, in a way as if the faulty sensor has been repaired. Without the digital twin, the TMR structure will not be able to output the correct value of the sensed parameter if one of the remaining two sensors experiences a transient or interment fault as the voting structure in the presence of only two working systems does not work.
Thus, the use of software-based digital twins as a duplicate for each physical sensor improves the reliability of the operation without overly increasing the overall system cost. 
The organization of the rest of the article is as follows: Section II details the proposed redundancy structure which is realized with triplicated sensors along with a digital twin for each of the sensors. The digital twin is based on the use of one of the popular time series forecasting algorithms as explained in Section II. Section III provides a detailed analysis of the effectiveness of this redundancy scheme including simulation results. Section IV gives an overview of the TMR Based Data Fusion scheme. Finally, Section V provides the conclusion and potential future work.

\section{Digital Twin Based Redundancy}

Figure 1 illustrates the organization of the proposed fault-tolerant IoT structure. The actual implementation envisions each critical IoT device to be triplicated and the output of each device is connected to an anomaly detector of the type discussed in [17]. This anomaly detector is based on the use of an AI/ML-based software model, and it detects mismatches among the IoT device outputs reliably and with a short reaction time. 
As explained earlier, the scheme relies on TMR. If any one of the three systems fails, the other two systems can correct and mask the fault. The general case of TMR is called N-modular redundancy, in which any positive number of replications of the same action is used. The number is typically taken to be at least three, so that error correction by majority vote can take place; it is also usually taken to be odd (N could be 5 or 7 or 9 etc.).
In our proposed scheme, which is shown in Figure 1, each physical sensor is paired with a software-based digital twin which takes over the function of the corresponding physical sensor when it malfunctions. The digital twin at that point continues to participate in the majority voting scheme, thus maintaining the integrity of the majority voting structure till the faulty physical device is either replaced or repaired.
The digital twin is less expensive than the actual physical IoT device since it does not need to have a “sensing” mechanism. It has a software model that continuously tracks the actual sensor readings and in the case of the corresponding sensor fault, the digital twin “interpolates” the future readings as close to the actual sensor readings based on the “Time Series Model” embedded in the digital twin.

\begin{figure}[h]
\centering
\includegraphics[scale=0.22]{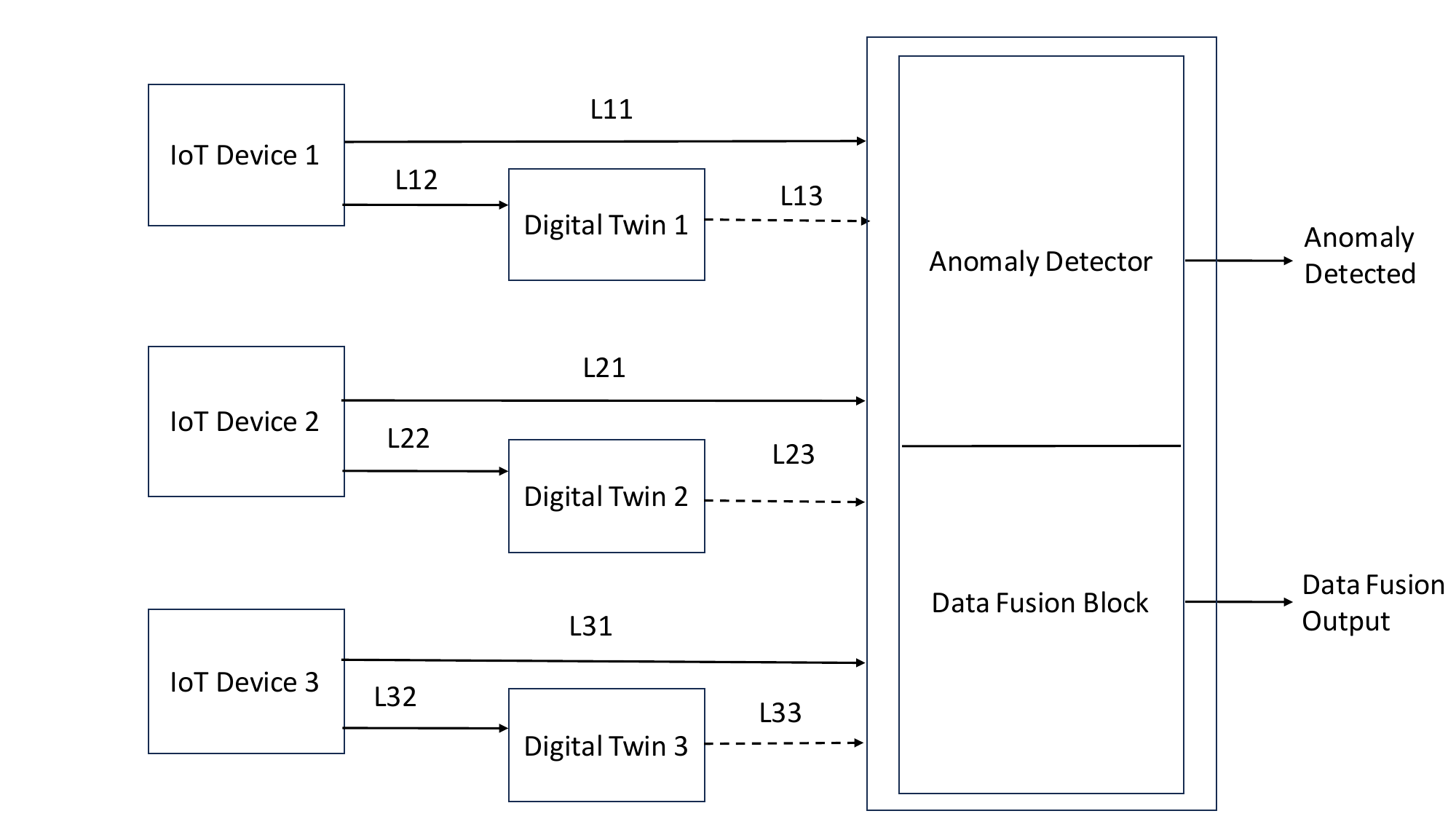}
\caption{Flow-chart of the proposed fault-tolerant IoT structure. We leverage shadow-IoT sensors acting as digital twins that are substituted for anomalous sensors to preserve the majority voting system in the case of a subsequent anomaly in one of the other sensors. The anomaly detector would also be able to detect once the shadow-IoT sensor deviates from the other true readings and drop it out. }
\label{Figure1}
\end{figure}

As indicated earlier, in the implementation of software-based digital twins, the capability to predict future values of sensed parameters is based on the use of time series forecasting algorithms [18-20]. There are many useful time series forecasting models discussed in the literature [18-20].  The simplest one is based on the concept of Kalman Filtering [21]. The reference [18] discusses several useful Time Series Forecasting algorithms that can help understand how historical data influences the future. This is done by looking at past data, defining the patterns, and producing short or long-term predictions.
In the proposed implementation in this paper, two of the popular time series forecasting algorithms are investigated for their applicability as a software-based digital twin. The algorithms investigated are: (a) Prophet [19] which is an open-source library developed by Facebook and designed for automatic forecasting of univariate time series data and (b) Temporal Fusion Transformer (TFT): a novel attention-based architecture developed by Google that combines high-performance multi-horizon forecasting with interpretable insights into temporal dynamics [20].
Prophet is based on an additive model where non-linear trends are fit with yearly, weekly, and daily seasonality, plus holiday effects. It also allows for modeling of the effects of changes in trend over time and includes a built-in procedure for detecting and handling outliers.
On the other hand, the Temporal Fusion Transformer (TFT) is an attention-based deep neural network that combines high-performance multi-horizon forecasting with interpretable insights into temporal dynamics.
While the Prophet is a simpler algorithm that is easier to use and interpret, the TFT, though a more complex algorithm, can provide more accurate predictions for multi-horizon forecasting problems. However, TFT requires more data and computational resources than Prophet. The choice between them depends on the specific problem and the desired trade-offs between interpretability and performance.

Referring to Figure 1, the fault-tolerant structure works as follows:

\begin{itemize}
\item Each IoT device is tracked by a digital twin based on either the Prophet or TFT model which forecasts time series data.
\item When an anomaly is detected, the IoT device triggering the anomaly is disconnected from the Anomaly Detector. Instead, the digital twin corresponding to this IoT sensor is connected to the Anomaly Detector.
\item If the malfunction of the defective IOT Device is of short duration (hours or few days), the digital twin of the IoT model “bridges” the gap through interpolation (forecasting) of future data based on the learning from the past data.
\item If the duration is long, the digital twin of the IoT Model will decay and diverge sufficiently from the other two IoT devices, triggering the anomaly detector again.
\item At that point, the output of the digital twin is disconnected from the Anomaly Detector and the system continues with two functioning IoT devices till the defective IoT device is either repaired or replaced.
\item If an anomaly is detected before the repair action, the usual practice in TMR structure is to randomly select one of the devices for use in measuring the parameter to be sensed.
\end{itemize}

In the next section, an analysis of the effectiveness of the implementation of the digital twin based on the Prophet algorithm and TFT algorithm is performed.

\section{Analysis of The Effectiveness of Two Implementations}

A time series forecasting model comprises 4 general components as follows [18]:
\begin{itemize}
\item {\bf Trend:} Increase or decrease in the series of data over a long period.
\item {\bf Seasonality:} Fluctuations in the pattern due to seasonal determinants over a period such as a day, week, month, or season.
\item {\bf Cyclical variations:} Occurs when data exhibit rises and falls at irregular intervals.
\item {\bf Random or irregular variations:} Instability due to random factors that do not repeat in the pattern.
\end{itemize}

In this section, the performance of the Prophet and TFT-based forecasting algorithms are compared using real-world sensor data [22].

\subsection{Performance of digital twin of the IoT implementation using Prophet}

The Prophet-based model is used as a digital twin in the first set of simulations which were run on the Intel Berkeley Data [22]. This data is collected from 54 sensors deployed in the Intel Berkeley Research lab between February 28th and April 5th, 2004 [22]. The sensors collected time-stamped topology information, along with humidity, temperature, light, and voltage values once every 31 seconds.

With a Prophet-based model used as a digital twin, when an anomaly is detected for a certain sensor, the Prophet model, trained on past ‘n’ sensor readings, is used as a substitute for the faulty sensor.

In this experiment, the model was trained on 6500 minutes of prior sensor readings from Intel Berkeley Data, which corresponds to 10000 total readings (because the readings are not exactly spaced out every minute). The training exercise took 7.0 seconds. This training model was used to predict a reading every minute for the following 3400 minutes, which corresponded to 5000 real data readings. It is to be noted that the Intel Berkeley Data is not temporally uniform.

\begin{figure*}[ht]
\centering
\includegraphics[scale=0.40]{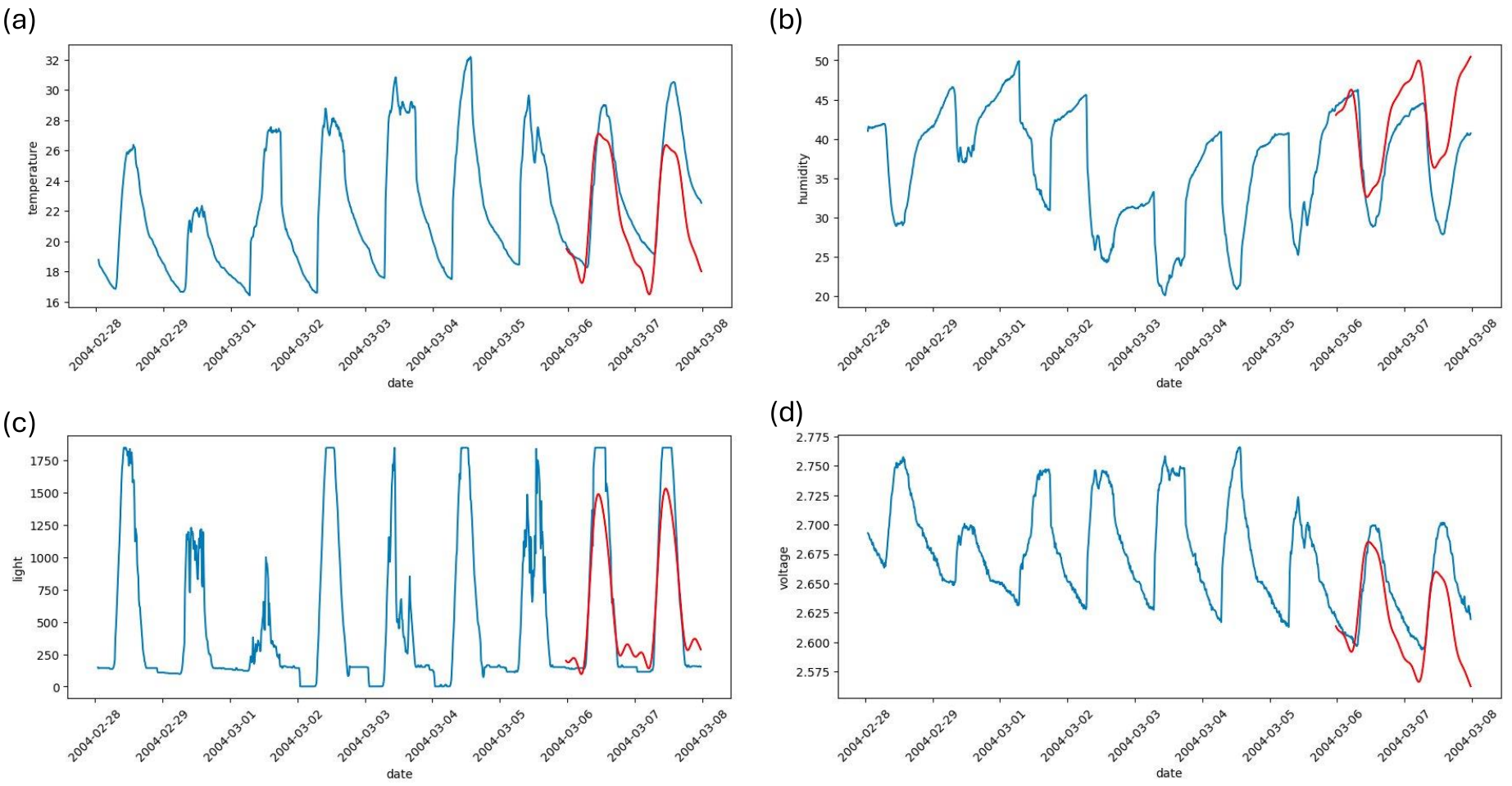}
\caption{Performance of Prophet model on the different sensor types. Shown in blue is the true sensor readings that the Prophet model was trained on, and in red is the digital twin generated by Prophet. (a) shows the predictions on the temperature sensor. Prophet tracks the sensor closely from the start of the prediction to the start of 2004-03-07. (b) shows the predictions on the humidity sensor. Prophet tracks the sensor closely from the start of the prediction to the middle of 2004-03-06 and preserves the underlying seasonality for the entire 2-day period.  (c) shows the predictions for the light sensor. Prophet tracks the sensor closely for the entire 2-day period. (d) shows the predictions on the voltage sensor. Prophet tracks the sensor closely until 2004-03-07}
\label{Figure2-New}
\end{figure*}

The full prediction time for the 3400 minutes was 19ms, which ended up being 5.6 us per timestep. In practice, the full range of future predictions could be done at once, to take advantage of vectorization. This leads to 7.019 total seconds of the sensor being inactive before the prediction model is started up. 

In the proposed implementation, the predictive model based on Prophet is continuously updated while the sensor is functioning correctly. This avoids the delay to 19ms of prediction time.

Please refer to Figure 2 for the performance of the Prophet Model. Shown in blue is the true sensor readings that the Prophet model was trained on, and in red is the digital twin generated by Prophet. The Figure 2 (a) shows the predictions on the temperature sensor. Prophet tracks the sensor closely from the start of the prediction to the start of 2004-03-07. The Figure 2 (b) shows the predictions on the humidity sensor. Prophet tracks the sensor closely from the start of the prediction to the middle of 2004-03-06 and preserves the underlying seasonality for the entire 2-day period. The Figure 2 (c) shows the predictions for the light sensor. Prophet tracks the sensor closely for the entire 2-day period. The Figure 2 (d) shows the predictions on the voltage sensor. Prophet tracks the sensor closely until 2004-03-07.

Based on the results depicted in Figure 2, the Prophet model as a substitute for the physical IoT device works quite well.

\subsection{Performance of digital twin of the IoT implementation using TFT}

As in Section III.A, the simulation experiments with TFT were also run using the Intel Berkeley Dataset [22]. 

The TFT model was trained to jointly output the temperature, humidity, voltage, and light sensor readings. The training dataset included readings from 5 sensor clusters at once, each cluster consisting of the temperature, humidity, voltage, and light sensor readings. The TFT model considers a history context of four days’ worth of good sensor readings and predicts the next two days’ readings. That means at the time of an anomaly, the TFT model would take as an input the past four days’ worth of good sensor readings before the occurrence of the anomaly and output the following two days’ readings.

Please refer to Figure 3 for the performance of the TFT Model. Shown in blue is the true sensor readings that the TFT model was trained on, and in red is the digital twin generated by TFT. The Figure 3 (a) shows the predictions on the temperature sensor. TFT tracks the actual sensor reading closely, but it starts to deviate from the true sensor readings a bit in the middle of 2004-03-04. The Figure 3 (b) shows the predictions on the humidity sensor. TFT tracks the sensor closely from the middle of 2004-03-03 till the end of the two-day prediction range. The Figure 3 (c) shows the predictions for the light sensor. TFT tracks the sensor closely for the full two-day prediction range. The Figure 3 (d) shows the predictions on the voltage sensor. The TFT struggles to learn the voltage data, and outputs values that deviate more than what TFT is able to do in the case of the light, humidity, and temperature sensor readings.

\begin{figure*}[ht]

\centering

\includegraphics[scale=0.40]{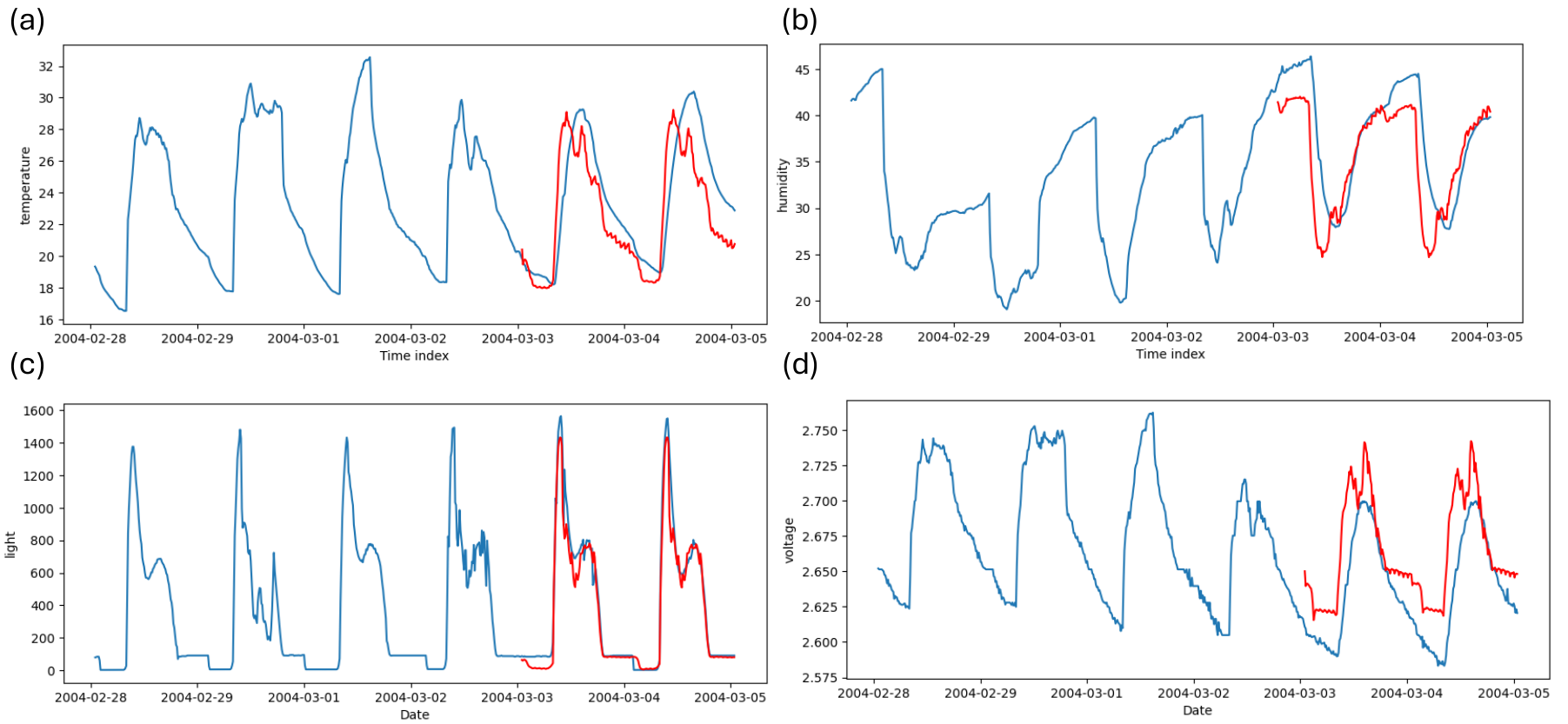}
\caption{Performance of TFT model on the different sensor types. Shown in blue is the true sensor readings that the TFT model was trained on, and in red is the digital twin generated by TFT. (a) shows the predictions on the temperature sensor. TFT tracks the sensor closely, but it starts to deviate from the true sensor readings a bit in the middle of 2004-03-04. (b) shows the predictions on the humidity sensor. TFT tracks the sensor closely from the middle of 2004-03-03 till the end of the two-day prediction range.  (c) shows the predictions for the light sensor. TFT tracks the sensor closely for the full two-day prediction range. TFT can predict the spikes in the light readings and the structure of the data as well. (d) shows the predictions on the voltage sensor. TFT. TFT struggles to learn the voltage data, and outputs values that deviate more than the other light, humidity, and temperature sensor readings.}
\label{Figure3-New}
\end{figure*}

Based on the results depicted in Figure 3, the TFT model as a substitute for the physical IoT device works quite well.

\subsection{Comparison of Prophet and TFT}

The Prophet model tracks the real sensors for temperature, humidity, light, and voltage very well for nearly one day. This is a long enough time for the real sensor to get repaired before the anomaly results in unacceptable consequences. The weakness of the Prophet model is that it needs to be trained for each sensor type and each sensor of the same type separately. However, this weakness is mitigated by the fact that the training time for the Prophet model is superfast, and the training resources required are minimal.

The TFT [20, 23] model tracks the real sensors for temperature, humidity, and light very well for two full days, but not for voltage. This is because the Intel Berkeley Dataset for voltage is very noisy and thus the fact that TFT does not track very well for voltage data is more of an artifact of the training data available. For clean data, even for voltage, the TFT model should work very well. In general, TFT mimics the real sensors for nearly two days, while the Prophet tracks the sensors for one day. Even though the TFT model performs better, it suffers from longer training time and requires more training resources. This weakness is somewhat mitigated by the fact that the TFT model does not need to be trained for each sensor type and each sensor of the same type. 

Even for hard faults, both the models work very well for one full day, with TFT having an edge that this model can be effective for even two full days. Since the use of the digital twin concept always maintains the integrity of the TMR structure, the proposed fault-tolerant approach is effective not only for hard faults but also for soft, intermittent, and transient faults.

\section{TMR Based Data Fusion}
Referring to Figure 1, there are two centralized blocks in the Fault-Tolerant IoT Systems being considered in this paper. The first one is the Anomaly Detector which was described in Section II. The second one is a TMR based data fusion block which collects all information from different sensors and computes the overall value of the parameter being measured by three individual sensors [24]. The output of the three physical sensors is processed by a majority-voting system [17] during data fusion process to produce a single output. If any one of the three systems fails, the other two systems can correct and mask the fault. The data fusion algorithm works as follows: 

\begin{enumerate}[label=\alph*)]
\item Assume that the three sensors send out readings of 5 previous time stamps individually to the data fusion block: 
\item Assuming, the readings are: Sensor-1: (7,8,8,7,6), Sensor-2 : (6,7,8,8,7), Sensor-3 : (7,6,7,8,8), the output based on Data Fusion would be:  (6.66, 7, 7.66, 7.66, 7), where each element of the output set is the average of the individual elements of the 3 sensors. The composite output would be the average of (6.66, 7, 7.66, 7.66, 7) which is equal to 7.20. This represents the best estimate for the parameter being read by the three sensors based on 5 previous readings of the same parameter by the three sensors. 
\item The scheme can deal with fault conditions very effectively. For instance, if the sensor 2 sustains a fault and outputs a value 0 instead of the correct sensor reading, the readings of the 3 sensors will be as follows: Sensor-1: ( 7,8, 8,7,6 ), Sensor-2: ( 6,0,0,8,7 ), Sensor-3: ( 7,6,7,8,8). Thus, the Sensor-2 has two missing numbers. The missing numbers are replaced by the composite value (the estimate of the last cycle) of the reading which was 7.20. If there is no record of a previous composite value or estimate, compute an estimate for this cycle using the procedure in (b) above. 
\item The Auto-Corrected reading based on estimate from step (b) would be: Sensor-1 ( 7,8,8,7,6), Sensor-2 ( 6, 7.20, 7.20, 8, 7), Sensor-3 ( 7,6,7,8,8). The TMR Output is computed as before and is (6.66, 7.07, 7.4, 7.66, 7). The new Composite Value is computed as before and is 7.16, which is the best estimate after the second cycle of reading. 
\item Instead of look-back being 5-time stamps, it could be any number N. N should be large enough so that the average is meaningful but should not be so large that it takes a long time before TMR-based fusion cycle is repeated.
\item In order to decide whether a reading in one of the sensors is erroneous when the sensor reading is not zero but another number, each element of a sensor reading value is compared with the Composite Value (which is the best estimate at any given point) and if the deviation is more than a “threshold”, the sensor reading is deemed erroneous and needs to be corrected. 
In that case, the erroneous reading is replaced with the last composite value. This algorithm detects and    corrects hard faults, soft faults, intermittent faults, and transient faults. 
\end{enumerate}

\section{Conclusion}

In this article, we present a novel redundancy scheme to realize a fault-tolerant IoT structure for application in high-reliability systems. The proposed fault-tolerant structure uses a centralized data fusion block and triplicated sensors, along with a software-based digital twin for each sensor. The digital twin is realized using a time series forecasting algorithm so that the digital twin can mimic the function of each sensor for some time. 

Two of the popular time series forecasting algorithms, Prophet and TFT, are investigated for their applicability as a software-based digital twin device. Based on simulations, both of the models have been found to be immensely suitable for the application intended in this paper. 
The use of software-based digital twins as a duplicate for each physical sensor improves the reliability of the operation without overly increasing the overall system cost. Because of the use of digital twin concept along with the proposed data fusion scheme, the TMR structure is able to output the correct value of the sensed parameter when one of the sensors sustains a hard fault and in addition one of the remaining two sensors experiences a soft, transient, or intermittent fault. 
In fact, the structure is able to output correct value of the sensed parameter under multiple fault conditions.

The future research investigations will include experiments on: a) longer prediction windows, b) different datasets, and c) effects of multiple simultaneous sensor failures.

\end{document}